\documentclass[a4paper,aps,prd,preprintnumbers,notitlepage,twocolumn,showpacs,superscriptaddress,nofootinbib]{revtex4-1}
\usepackage{graphics,graphicx,epsfig}
\usepackage{amsfonts,amsmath,amssymb}
\usepackage{amsmath}
\usepackage{amsfonts}
\usepackage{amssymb}
\usepackage{color}
\usepackage{graphicx}
\usepackage{float}

\definecolor{darkgreen}{rgb}{0,0.35,0}

\begin{document}

\title{Chemical potential driven phase transition of black holes in Anti-de Sitter space}
\author{Mario Galante}
\email{mario-at-df.uba.ar}
\affiliation{Departamento de F\'{\i}sica, Universidad de Buenos Aires and IFIBA-CONICET,
Ciudad Universitaria, Pabell\'on I, 1428, Buenos Aires, Argentina.}
\author{Gaston Giribet}
\email{gaston-at-df.uba.ar}
\affiliation{Physique Th\'eorique et Math\'ematique, Universit\'e Libre de Bruxelles and International Solvay Institutes,
Campus Plaine C.P. 231, B-1050 Bruxelles, Belgium.}
\affiliation{Departamento de F\'{\i}sica, Universidad de Buenos Aires and IFIBA-CONICET,
Ciudad Universitaria, Pabell\'on I, 1428, Buenos Aires, Argentina.}
\affiliation{Instituto de F\'{\i}sica, Pontificia Universidad Cat\'{o}lica de Valpara%
\'{\i}so, Casilla 4059, Valpara\'{\i}so, Chile.}
\author{Andr\'es Goya}
\email{af.goya-at-df.uba.ar}
\affiliation{Departamento de F\'{\i}sica, Universidad de Buenos Aires and IFIBA-CONICET,
Ciudad Universitaria, Pabell\'on I, 1428, Buenos Aires, Argentina.}
\author{Julio Oliva}
\email{julioolivazapata-at-gmail.com}
\affiliation{Departamento de F\'{\i}sica, Universidad de Concepci\'on, Casilla 160-C, Concepci\'on, Chile.}

\begin{abstract}
Einstein-Maxwell theory conformally coupled to a scalar field in $D$ dimensions may exhibit a phase transition type instability at low temperature. We find an explicit solution that represents a charged asymptotically Anti-de Sitter black hole with a scalar field profile that is regular everywhere outside and on the horizon. This provides a tractable model to study the phase transition of hairy black holes in Anti-de Sitter space in which the backreaction on the geometry can be solved analytically.
\end{abstract}

\maketitle

\section{Introduction}   

Asymptotically Anti-de Sitter (AdS) black holes represent holographic duals of strongly coupled conformal field theories at finite temperature, and the phase transition phenomena in those backgrounds admit an interpretation as such in the dual boundary theory too. The interpretation of Hawking-Page phase transition \cite{HP} of 5-dimensional AdS black holes as the phenomenon dual to the confinement/deconfinement transition in the quark-gluon plasma in 4-dimensional Yang-Mills theory \cite{Witten}, and the interpretation of the scalar field condensation around 4-dimensional charged AdS black holes as holographic models of superconductors \cite{HHH, Gubser} are two archetypical examples of this general idea. The relevance of AdS black holes in the context of AdS/CFT correspondence \cite{Maldacena} motivates the seek of tractable models in which the phase transition of higher-dimensional AdS black holes coupled to matter fields can be solved explicitly including backreaction. This is because having such a model would enable one to describe these critical phenomena and analogs analytically. Unfortunately, coming up with such a model is difficult and the catalog of scenarios in which this is under control is actually scant. Nevertheless, surprisingly, a model in which phase transitions of AdS black holes in arbitrary dimension $D$ does admit an analytic solution is given by general relativity conformally coupled to a real scalar field \cite{Nos2}. Such hairy black hole solutions in arbitrary $D>4$ were recently found in Ref. \cite{Nos1} (see  \cite{Troncoso1, Troncoso2, Anabalon} for lower-dimensional analogs), and it was argued in \cite{Nos2} that they serve as models to study the condensation of scalar hair around an asymptotically AdS black hole including the backreaction of the field on the metric: Above certain critical temperature, the theory undergoes a phase transition whose endpoint turns out to be an asymptotically AdS black hole with a scalar field configuration that is regular everywhere outside and on the horizon \cite{Nos2}. However, the example considered in \cite{Nos2} failed to model the condensation of the scalar field at low temperature, and it rather exhibited such a transition only at high temperature. This motivates us to try to modify the ultra-violet behavior of the solution of \cite{Nos1, Nos2} in such a way of inducing a phase transition at low temperature. A natural idea to do so is to charge the hairy black hole under a gauge field: In dimension $D>4$, this would yield a matter-energy contribution that, at short distance, would dominate over the conformally coupled matter and probably reinforce the backreaction in such a way that a new phase transition actually takes place. As we will see, this intuition turns out to be correct: The simplest scenario to achieve scalar field condensation at low temperature turns out to be coupling the theory of \cite{Nos0, Nos1} to a $U(1)$ gauge field in $D=5$ dimensions. We will see that the inclusion of charge to the hairy AdS black holes of \cite{Nos1, Nos2} reinforces the backreaction at short distances in such a way that it triggers a new type of instability at low temperature. The inclusion of electric charge is also natural from the holographic point of view: Within the context of AdS/CFT, the inclusion of a $U(1)$ field in the bulk corresponds to the inclusion of finite chemical potential. 

\section{The theory}

Consider Einstein-Maxwell theory conformally coupled to scalar field theory matter in 5 dimensions. The action of the theory is given by
\begin{equation}\label{actionEMS}
{\mathcal I} =\frac{1}{\kappa}\int d^5 x \sqrt{-g}\left( R-2\Lambda-\frac{1}{4}F^2+\kappa\mathcal{L}_m(\phi,\nabla\phi)\right) ,
\end{equation}
where $\kappa =16\pi G$, with $G $ being the 5-dimensional Newton constant. We will consider $\Lambda<0$, although solutions with the other sign of the 
cosmological constant are permitted too \cite{Nos1}. Apart from the cosmological Einstein-Hilbert and Maxwell terms, the theory includes a conformal field theory contribution: The Lagrangian density $\mathcal{L}_m(\phi,\nabla\phi)$ corresponds to the most general action for a real scalar 
field conformally coupled to gravity that results in second order field equations. The explicit construction of such Lagrangian in $D$ dimensions has been done 
in \cite{Nos0}, and it can be conveniently expressed in terms of a rank-4 tensor defined as
\begin{align}
S_{\mu\nu}^{\quad\gamma\delta}  &  =\phi^{2}R_{\mu\nu}^{\quad\gamma\delta
}-12\delta_{\lbrack\mu}^{[\gamma}\delta_{\nu]}^{\delta]}\nabla_{\rho}%
\phi\nabla^{\rho}\phi-\nonumber\\
&  48\phi\delta_{\lbrack\mu}^{[\gamma}\nabla_{\nu]}\nabla^{\delta]}%
\phi+18\delta_{\lbrack\mu}^{[\gamma}\nabla_{\nu]}\phi\nabla^{\delta]}%
\phi . \label{Sij}%
\end{align}

This tensor can be seen to transforms covariantly under local Weyl rescaling $g_{\mu\nu}\rightarrow\Omega^{2}g_{\mu\nu},\,\phi\rightarrow\Omega
^{-1/3}\phi$. In fact, one finds that under such transformation tensor (\ref{Sij}) transforms\footnote{In comparing with \cite{Nos0} one has to take into account that here we are considering the scaling dimension $s=-1/3$.} as $S_{\mu\nu}^{\ \ \gamma\sigma} \rightarrow\Omega
^{-8/3} S_{\mu\nu}^{\ \ \gamma\sigma}$.

In terms of tensor (\ref{Sij}), the $D$-dimensional version of the Lagrangian matter $\mathcal{L}_m(\phi,\nabla\phi)$ takes the form
\begin{align}\label{matter}
\mathcal{L}_m (\phi,\nabla\phi)  &  =\sum_{k=0}^{\left[  \frac{D-1}{2}\right]
}b_{k}\frac{k!}{2^{k}}\ \phi^{3D-8k}\delta_{\lbrack\alpha_{1}}^{\mu_{1}}%
\delta_{\beta_{1}}^{\nu_{1}} ...  \delta_{\alpha_{k}}^{\mu_{k}}\delta_{\beta
	_{k}]}^{\nu_{k}}\ \times\nonumber\\
&  \qquad\times S_{\quad\mu_{1}\nu_{1}}^{\alpha_{1}\beta_{1}} \ ... \ S_{\quad
	\mu_{k}\nu_{k}}^{\alpha_{k}\beta_{k}}\,,
\end{align}
where the symbol $[n]$ stands for the integer part of $n$, and $b_k$ are real arbitrary coupling constants. Tensor (\ref{Sij}) has the symmetry of the Riemann tensor under indices permutation. In fact, it can be shown to represent a curvature tensor too. Therefore, Lagrangian (\ref{matter}) is of the Lovelock type and is the natural higher-dimensional extension of the conformally coupled theory $\mathcal{L}=-(\nabla \varphi)^2-(1/6)R\varphi^2-\lambda\varphi^4$ in 4 dimensions \cite{Nos0, Nos1}. 

In this paper, and mainly because of its relevance for holography, we will be concerned with the special case $D=5$. Nevertheless, we emphasize that the solutions reported here, and in particular the electrically charged hairy solution, are straightforwardly generalized to arbitrary values of $D$. In the case $D=5$, (\ref{matter}) reads
\begin{eqnarray}
\mathcal{L}_m (\phi,\nabla\phi) &=& b_{0}\phi^{15}+
b_{1}\phi^{7} S_{\mu \nu }^{\ \ \mu \nu } +b_{2}\phi^{-1} (S_{\mu \gamma }^{\ \ \mu \gamma }S_{\nu \delta }^{\ \ \nu \delta }-\nonumber \\
&-& 4 S_{\mu \gamma }^{\ \ \nu \gamma } S_{\nu \delta }^{\ \ \mu \delta }+S_{\mu\nu}^{\ \ \gamma\delta}S^{\mu\nu}_{\ \ \gamma\delta}) .
\label{np}%
\end{eqnarray}

Notice that the fact that tensor (\ref{Sij}) is quadratic in the field $\phi $ prevents the couplings from developing singularities in the limit $\phi\rightarrow 0$. This observation is relevant because it enables one to consider the non-hairy black hole solutions ({\it{i.e.}} $\phi =0$) as part of the ensemble.

The equations of motion coming from the action above are the Einstein equations
\begin{equation}
R_{\mu\nu}-\frac{1}{2}Rg_{\mu\nu}+\Lambda g_{\mu\nu}=\kappa \ ^{(1)}T_{\mu\nu
}+ {} ^{(2)}T_{\mu\nu
} , \ \label{GT}%
\end{equation}
with the energy-momentum tensor contributions%
\begin{align}
^{(1)}T_{\mu}^{\nu}  &  =\sum_{k=0}^{\left[  \frac{D-1}{2}\right]  }\frac{k!b_{k}%
}{2^{k+1}}\phi^{3D-8k}\delta_{\lbrack\mu}^{\nu}\delta_{\rho_{1}}^{\lambda_{1}%
}...\delta_{\rho_{2k}]}^{\lambda_{2k}}\times\nonumber\\
&  \qquad\times\ S_{\ \ \ \ \lambda_{1}\lambda_{2}}^{\rho_{1}\rho_{2}%
} \ ... \ S_{\ \ \ \ \lambda_{2k-1}\lambda_{2k}}^{\rho_{2k-1}\rho_{2k}}\ ,
\label{TTT}\nonumber\\ 
^{(2)}T_{\mu}^{\nu} & = F_{\mu\rho}F^{\nu \rho }-\frac{1}{4}F_{\lambda \rho}F^{\lambda \rho}\delta_{\mu}^{\nu};
\end{align}
together with the Maxwell equations
\begin{align}
\nabla_{\mu}F^{\mu\nu} =0 \ , \ \ \ \nabla_{[\mu}F_{\nu\rho]} =0;
\end{align}
and the Horndeski type equation for the scalar field, namely
\begin{align}
0  &  =\sum_{k=0}^{\left[  \frac{D-1}{2}\right]  }\frac{\left(  D-2k\right)
	k!b_{k}}{2^{k}}\phi^{3D-8k-1}\delta_{\lbrack\alpha_{1}}^{\mu_{1}}\delta
_{\beta_{1}}^{\nu_{1}}...\delta_{\alpha_{k}}^{\mu_{k}}\delta_{\beta_{k}]}%
^{\nu_{k}}\times\nonumber\\
&  \qquad\times\ S_{\quad\mu_{1}\nu_{1}}^{\alpha_{1}\beta_{1}} \ ... \ S_{\quad
	\mu_{k}\nu_{k}}^{\alpha_{k}\beta_{k}}. \label{eqfieldarb}%
\end{align}

It is important to remark that the trace of the contribution $^{(1)}T_{\mu\nu}$ vanishes on-shell due to the conformal invariance of $\mathcal{L}_m (\phi,\nabla\phi)$.

\section{Charged black holes}

Analytic black hole solutions to the theory defined by the field equations (\ref{GT})-(\ref{eqfieldarb}) in $D$ dimensions and with non-vanishing $\phi $ have been found in Ref. \cite{Nos1}. These represent electrically uncharged black holes with a scalar hair that is regular everywhere outside and on the horizon. Here, we will see that, remarkably, the results of \cite{Nos1} can be extended to the case of electrically charged solutions. 

To see this explicitly, consider the static spherically symmetric ansatz
\begin{equation}
ds^{2}=-N^{2}(r)f(r)\ dt^{2}+\frac{dr^{2}}{f(r)}+r^{2}d\Omega_{3}^{2} .
\label{g1}%
\end{equation}

It is possible to see that field equations (\ref{GT})-(\ref{eqfieldarb}) in $D=5$ dimensions admit an exact solution of the form (\ref{g1}), with
\begin{equation}
f(r)=1-\frac{m}{r^{2}}-\frac{q}{r^{3}}+\frac{e^2}{r^4}-\frac{\Lambda}{6}r^{2},\qquad
N^{2}(r)=1, \label{g2}%
\end{equation}
where $d\Omega^2_3$ is the metric of the unit 3-sphere (of volume $2\pi^2$), and where $e$ and $m$ are two integration constants associated to the 
mass and the electric charges respectively (see (\ref{GasM})-(\ref{GasQ}) below). In (\ref{g2}), $q$ is given in terms of the coupling constants by
\begin{equation}
q=\frac{64\pi G}{5}\varepsilon b_1  \left(  -\frac{18}{5}\frac{b_{1}}{b_{0}}\right)  ^{3/2} , \label{conditions}%
\end{equation}
with $\epsilon=-1,0,+1$. For the black hole solution (\ref{g1})-(\ref{g2}) to exist, the coupling constants have to satisfy an additional constraint \cite{Nos1}, namely
\begin{equation}
10b_{0}b_{2}=9b_{1}^{2}. \label{conditionsbis}
\end{equation}
That is, $q$ is a discrete parameter that can take only three different values, namely $q=0,\pm|q|$.

The scalar field configuration, on the other hand, is given by\footnote{Despite the weakened falloff of the scalar field, which goes like $\phi \sim 1/r^{1/3}$, the non-minimal couplings with the curvature prevent the mass of the solution from receiving contributions from the field.}
\begin{eqnarray}
\phi(r)=\dfrac{n}{r^{1/3}} , \ \ \ \ n=\varepsilon
\left(  -\frac{18}{5}\frac{b_{1}}{b_{0}}\right)  ^{1/6}. \label{phi}%
\end{eqnarray}

As $q$, the absolute value of $n$ is fixed in terms of the coupling constants $b_k$ and can take only three values, namely $n=0,\pm|n|$. We see from (\ref{g2}) that $q$ can be thought of as the {\it charge} of the black hole under the scalar field. Nevertheless, it is worth emphasizing that $q$ is neither a charge nor hair in the usual sense, as it may only take discrete values, with its absolute value being fixed in terms of the coupling constants. In the case $q=0$ the solution reduces to the Reissner-Nordstr{\o}m solution in 5-dimensional AdS space. In all cases, the solution asymptotes to AdS at large distance.

The solution for the gauge field takes the electric form
\begin{equation}
A_{\mu}=\sqrt{3}\frac{e}{r^2} \delta_{\mu}^{0}, \label{gasA}
\end{equation}
with $F_{\mu\nu}=\partial_{\mu }A_{\nu}-\partial_{\nu }A_{\mu}$.

The existence of the black hole solution (\ref{g1})-(\ref{gasA}) in $D$ dimensions is remarkably in its own right \cite{Martinez}. The full solution is characterized by three parameters, namely $m$, $e$, and $\varepsilon $. While the electric charge of the black hole is given by $e$, its charge under the scalar field is given by $q $. Unlike the former, the net value of the latter charge is fixed in terms of the coupling constants. However, since in particular it can be zero, it has to be regarded as a discrete charge parameter that controls the intensity of the scalar field, which is given by $n$.

Apart from the Planck length $\ell_P = G^{1/3}$, the problem has three different characteristic length scales that are relevant, namely $|\Lambda |^{-1/2}$, $|e|^{1/2}$, and $|q|^{1/3}$. The physical interpretation of these scales is what ultimately enables us to identify a region of the parameter space that yields the desirable critical phenomenon. While $\ell \equiv (-\Lambda /6)^{-1/2}$ controls the physics of AdS black holes at large distances, the interplay between the parameters $|e|^{1/2}$ and $|q|^{1/3}$ is responsible for the critical behavior of small black holes.

For some ranges of the parameters, the solution represents a black hole. Due to the form of the metric function (\ref{g2}), these black holes behave as Reissner-Nordstr{\o}m black holes in AdS at large distance. However, provided $q\neq 0$, their behavior is qualitatively different from the general relativity solutions at short distances. For instance, as observed in \cite{Nos2} for the uncharged ({\it i.e.} $e=0$) black holes, the case with $q< 0$ can present Hawking temperature arbitrarily low in AdS space and positive specific heat even for small black holes. We will see here that, in the case of electrically charged solutions, this is also the case for solutions with $q>0$. In \cite{Nos2}, the solutions with $q>0$ were regarded as pathological due to unphysical behaviors they exhibit at very short distance, such as negative mass configurations. This was physically interpreted in \cite{Nos1} as the contribution $^{(1)}T_{\mu \nu}$ in the stress-tensor to violate the energy conditions when $q$ is positive. However, here we will show that when the solution is electrically charged such pathologies disappear, even for small charge values. This is due to the new contribution $^{(2)}T_{\mu \nu}$ in the stress-tensor, which is in $D>4$ dimensions is dominant for black holes with small horizons.

The horizons of the black hole (\ref{g1})-(\ref{g2}) are given by the positive zeroes of the equation $f(r)=0$; that is, the positive roots of the polynomial 
\begin{equation}
r_+^6 - \ell^2 r_+^4 + {m}\ell^2 r_+^2 + {q}\ell^2 r_+ -{e^2}\ell^2  = 0 , \label{GasMasa}
\end{equation}
where $\ell^2 =6/|{\Lambda }|$. As for the Reissner-Nordstr{\o}m black hole in AdS space, the solution above may present inner and outer horizons. The location of these horizons, denoted $r=r_-$ and $r=r_+$ respectively, depend on $q$. Two horizons may exist both for positive and negative values of $q$. The extremal solution corresponds to the case where both horizons coincide, $r_{\text{ext}}\equiv r_-=r_+$; this happens when
\begin{equation}
r_{\text{ext}}^{6}+\frac{\ell^{2}}{2}r_{\text{ext}}^{4}
+\frac{q\ell^{2}}{4}r_{\text{ext}}-\frac{e^2\ell^2}{2 }=0.
\label{W}%
\end{equation}

To compute the conserved charges associated to the black hole solution above one may resort to the Regge-Teitelboim approach in the minisuperspace. This amounts to consider the ansatz (\ref{g1}) and vary the action functional ${\mathcal I}$, integrated in a finite time interval $(t,t+\Delta t)$, with respect to the shift function $N(r)$ and the function $f(r)$. This yields the value
\begin{eqnarray}
M = \frac{3\pi}{8G}m - M_0 \label{GasM}
\end{eqnarray}
for the mass, and the value 
\begin{eqnarray}
Q = -\frac{\sqrt{3}\pi}{8G}e - Q_0, \label{GasQ}
\end{eqnarray}
for the electric charge, with $M_0$ and $Q_0$ are two integration constants. These results for the conserved charges are re-obtained in the Euclidean action formalism where the constants $M_0$ and $Q_0$ are found to be zero. This corresponds to choose AdS$_5$ space as the zero-mass configuration. This seems to be a reasonable choice. Nevertheless, let us point out that, in some cases, assigning $M=0$ to the AdS configuration is not necessary the correct answer. For instance, in three-dimensional gravity, AdS$_3$ space has actually mass given by $M=-1/(8G)$; this is due to the existence of a gap in the black hole spectrum. One can also find natural to assign the zero mass value to the extremal configuration in some cases.

\section{Black hole thermodynamics}

The Hawking temperature associated to the black hole (\ref{g1})-(\ref{g2}) is given by
\begin{equation}
T=\dfrac{1}{\pi\ell^{2}r_{+}^{4}}\left(-\frac{e^2 \ell^2}{2 r_{+}}+  \frac{q \ell^{2}}{4}%
+\frac{\ell^{2}}{2}r_{+}^{3}+r_{+}^{5}\right)  \, . \label{T}%
\end{equation}

This value vanishes in the extremal case. In the limit $q\to 0$, this reproduces the result for the AdS Reissner-Nordstr{\o}m black hole. Notice that, at certain scales, the contribution that depends on $q$ competes with the one that depends on the electric charge $e$. In the case $e=0$, the short distance behavior is governed by the $q$-dependent term in (\ref{T}). Because of that, the temperature of small black holes with $q>0=e$ diverges. In contrast, when the electric charge is non-vanishing, the black holes with $q>0$ shares the qualitative behavior of those with $q<0$ at short distance, in the sense that the specific heat changes its sign for small black holes. For the same reason, one observes from (\ref{GasMasa}) that the electrically charged solutions with $q>0$ do not exhibit the negative mass spectrum pathology diagnosed in \cite{Nos2} for the case $e=0<q$.

The result for the temperature (\ref{T}) enables one to consider the Euclidean action formalism. Here, we are concerned with the ensemble with fixed charge, and analyze the system at fixed temperature and volume (the gravitational potential of AdS acts as finite-volume reservoir). Then, we are interested in the Helmholtz free energy $F=M-TS$. In the semi-classical approximation this is given by
\begin{equation}
F = -\beta^{-1}\log Z \approx \beta^{-1}\mathcal{I}_{E}, \label{F}%
\end{equation}
with the period in the Euclidean time given by $\beta=1/T$. The result reads\footnote{A similar computation has been done in Ref. \cite{Nos2}. We refer to that paper for the details.}
\begin{eqnarray}
F&=&-\frac{5 \pi  e^2 q}{8 G r_+^5}+\frac{5 \pi  q^2}{16 G r_+^4}+
\frac{5 \pi  e^2}{8 G r_+^2}+\frac{\pi  q}{8 G r_+}+ \nonumber \\&+&
\frac{5 \pi  q r_+}{4 G \ell^2}+\frac{\pi  r_+^2}{8 G}-
\frac{\pi  r_+^4}{8 G \ell^2}  . \label{FfromI}
\end{eqnarray}

It is worth mentioning that the Euclidean action computation of the free energy $F$ involves a regularization of the infrared divergence, which can be achieved by background substraction in the standard way; that is, regularizing with respect to the thermal AdS$_5$ configuration (see for instance \cite{Nos2}).

In the Euclidean action formalism, the mass is obtained by
\begin{equation}
M=\frac{\partial\mathcal{I}_{E}}{\partial\beta}=\dfrac{3\pi}{8G}\left( \dfrac{e^2}{r_{+}^2} -\dfrac{q}{r_{+}}%
+r_{+}^{2}+\dfrac{r_{+}^{4}}{\ell^{2}}\right) , \label{Me}%
\end{equation}
which agrees with (\ref{GasM}) for $M_0=0$. Analogously, the entropy is given by
\begin{equation}
S=\beta\frac{\partial\mathcal{I}_{E}}{\partial\beta}-\mathcal{I}_{E}=\dfrac{\pi^{2}}{2G}r_{+}^{3}-\dfrac{5\pi^2 }{4G}q\, , \label{S}%
\end{equation}
which can be written as
\begin{equation}
S=\dfrac{A}{4G} - \dfrac{A_0}{4G}, \label{A}
\end{equation}
with $A_0=5\pi^{2}q$. While the first term on the right hand side realizes the Bekenstein-Hawking area law, the second term is an additive contribution that only depends on $\varepsilon $ and so it only contributes to the entropy of solutions with $\phi \neq 0$. The existence of a additive contribution to the area law in (\ref{A}) is not a surprise due to the higher-curvature terms couplings in the action, and this contribution will be of crucial importance in our discussion. The fact that the second term on the right hand side of (\ref{A}) does not depend on $r_+$ is a direct consequence of the fact that the scalar matter contribution is conformal invariant.

\section{Low temperature phase transition}

From the result for $F$ one can compare the free energy associated to the different AdS black hole solutions in relation to that of the thermal AdS space. We are interested in the problem at fixed temperature and fixed effective volume, the latter being provided by AdS space; and we also consider the problem at fixed electric charge. Therefore, negative values of (\ref{FfromI}) correspond to thermodynamically favored configurations relative to thermal AdS.  

The black hole solution presented here exhibits a rich parameter space, with two continuous variables ({\it i.e.} $m$, $e$) and one discrete variable ({\it i.e.} $\varepsilon$), which seems difficult to explore exhaustively. However, the physical interpretation of the characteristic length scales involved in the problem ({\it i.e.} $|\Lambda |^{-1/2}$, $|e|^{1/2}$, and $|q|^{1/3}$) permits to make an educated search and identify regions of the parameter space in which the critical phenomenon at low temperature actually takes place. See for instance the figure below, corresponding to the values of the parameters specified in the epigraph: As in \cite{Nos1}, one verifies the existence of a phase transition at high temperature, which yields a hairy black hole solution with $q<0$ provided $T$ is higher than certain critical value. This is seen from the fact that, at high temperature, the lowest free energy corresponds to the solution with negative values of $q$. 
\begin{figure}[H]
\begin{centering}
	\includegraphics[scale=1.0]{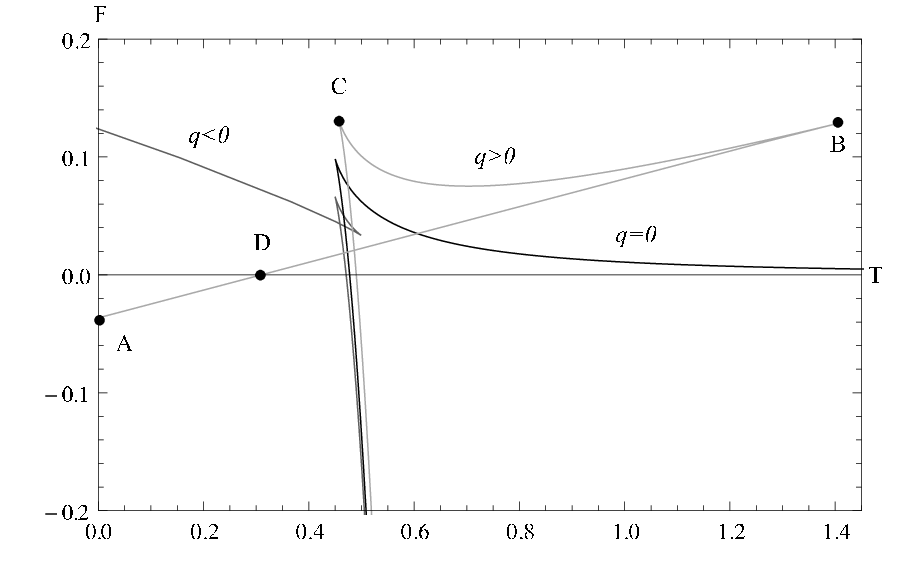}
	\caption{Free energy ($F$) as a function of the temperature ($T$) corresponding to the values $e= 0.025$, $q= \pm 0.9 \times 10^{-2}$, $\ell = 1$, $G= 
1$. At high $T$ the dominant configuration corresponds to the black hole with $q<0$. At low $T$, a branch with negative entropy configurations appears.}
	\label{phase}
	\end{centering}
\end{figure}

Phase transitions from thermal AdS to black hole configurations correspond to the points at which the curve of $F$ as a function of $T$ intersects the axis $F=0$. The novel feature that the electrically charged solution presented here exhibits is that, in addition to the high $T$ phase transition, the system exhibits a crossing point at low temperature. This is illustrated by the point D in the figure above. This is due to the fact that small black holes with $e\neq 0 <q$ change the sign of their specific heat at a critical size governed by the scales $q^{1/3}$ and $|e|^{1/2}$. This is expressed by the cusp at the point B in the figure (the line AB and the cusp at B disappear when $e=0$). The configurations on the line AB correspond to solutions with negative entropy\footnote{We thank Andr\'es Gomberoff for pointing out this feature to us.}. This strange feature occurs in higher-curvature theories \cite{Cvetic,Canfora}, and it is due to the contribution of the second term on the right hand side of (\ref{A}). In \cite{Cvetic} the existence of negative entropy configurations was interpreted as a type of instability. Notice that the term of the free energy (\ref{FfromI}) that dominates at short scales goes like $\sim -e^2q/r_+^5$. Therefore, provided the electric charge is non zero, the thermodynamically favored configuration at low temperature is a hairy black hole with positive $q$.

In summary, the coupling of the hairy black hole solutions to a $U(1)$ gauge field reinforces the backreaction at short distances in such a way that it triggers a new phase transition type instability at low temperature, and this suggests that the endpoint of the transition corresponds to a black hole in AdS space with non-vanishing scalar field configuration. This provides a tractable model to study the phase transition of hairy black holes in AdS space in which the backreaction on the geometry can be solved analytically.


\[
\]

The work of G.G. was partially funded by FNRS-Belgium (convention FRFC PDR T.1025.14 and convention IISN 
4.4503.15), by the Communaut\'{e} Fran\c{c}aise de Belgique through the ARC program and by a donation from the Solvay family. It was also supported by grants 
PIP0595/13 and UBACyT 20020120100154BA, from CONICET and UBA of Argentina. The work of J.O. is supported by FONDECYT grant 1141073 and by the Newton-Picarte Grant DPI20140053.

\end{document}